\documentclass[graybox]{svmult}

\usepackage{type1cm}        
\usepackage{makeidx}         
\usepackage{graphicx}        
\usepackage{multicol}        
\usepackage[bottom]{footmisc}

\usepackage{newtxtext}       %
\usepackage{newtxmath}       

\makeindex             

\begin{document}

\title*{Black holes: \\ on the universality of the Kerr hypothesis}
\author{Carlos A. R. Herdeiro}
\institute{Carlos A. R. Herdeiro\at Departamento de Matem\'atica da Universidade de Aveiro, Campus de Santiago,
 3810-193 Aveiro, Portugal, \email{herdeiro@ua.pt}}
%
%
\maketitle

\abstract*{To what extent are \textit{all} astrophysical, dark,  compact objects both black holes (BHs) and described by the Kerr geometry? We embark on the exercise of defying the \textit{universality} of this remarkable idea, often called the "Kerr hypothesis". After establishing its rationale and timeliness, we define a minimal set of reasonability criteria for alternative models of dark compact objects. Then, as proof of principle, we discuss concrete, dynamically robust non-Kerr BHs and horizonless imitators, that 1) pass the basic theoretical, and in particular dynamical, tests, 2) match (some of the) state of the art astrophysical observables and 3) only emerge at some (macroscopic) scales. These examples illustrate  how the universality (at all macroscopic scales) of the Kerr hypothesis can be challenged.}

\abstract{To what extent are \textit{all} astrophysical, dark,  compact objects both black holes (BHs) and described by the Kerr geometry? We embark on the exercise of defying the \textit{universality} of this remarkable idea, often called the "Kerr hypothesis". After establishing its rationale and timeliness, we define a minimal set of reasonability criteria for alternative models of dark compact objects. Then, as proof of principle, we discuss concrete, dynamically robust non-Kerr BHs and horizonless imitators, that 1) pass the basic theoretical, and in particular dynamical, tests, 2) match (some of the) state of the art astrophysical observables and 3) only emerge at some (macroscopic) scales. These examples illustrate  how the universality (at all macroscopic scales) of the Kerr hypothesis can be challenged.}

\section{The Kerr hypothesis}
\label{sec:1}
The elegant Kerr metric~\cite{Kerr:1963ud,Boyer:1966qh},
\begin{equation}
    ds^2=-\frac{\Delta}{\Sigma}\left(dt-a\sin^2\theta d\phi\right)^2+\frac{\Sigma}{\Delta}dr^2+\Sigma d\theta^2+\frac{\sin^2\theta}{\Sigma}[adt-(r^2+a^2)d\phi]^2 \ ,
    \label{kerrm}
\end{equation}
where $\Sigma\equiv r^2+a^2\cos^2\theta$, $\Delta\equiv r^2-2Mr+a^2$, is the currently accepted model to describe the phenomenology of \textit{all} astrophysical black hole (BH) candidates.  Since~\eqref{kerrm} is only described by two macroscopic parameters, the mass $M$ and angular momentum $Ma$ of the BH, this is clearly an economical scenario. The very same (almost featureless) theoretical model describes astrophysical objects ranging, at least, 10 orders of magnitude in mass. There is evidence for BHs in the stellar mass range, from $\sim$ $M_\odot$ to $\sim$100 $M_\odot$, obtained from X-ray binaries~\cite{Narayan:2013gca} and gravitational wave (GW) detections~\cite{LIGOScientific:2018mvr,LIGOScientific:2020ibl,LIGOScientific:2021djp} and in the supermassive range, from $\sim 10^5$ to $10^{10}$ $M_\odot$, as radio sources~\cite{Narayan:2013gca}, from Very Large Base Line Interferometry~\cite{EventHorizonTelescope:2019dse} and infrared observations in our own galactic centre~\cite{Gillessen:2008qv,Ghez:2008ms}. According to the "Kerr hypothesis" these BHs have exactly the same spacetime structure, simply rescaled by the mass, and with only one extra macroscopic degree of freedom, their spin. If true, this is remarkable. 

The goal of this article is to discuss, with concrete proof of concept models, the possibility if (and under which circumstances) astrophysical BH candidates could be described by something else rather than the Kerr geometry in some \textit{range of (macroscopic) scales},\footnote{It is widely accepted that sufficiently microscopic BHs will require quantum gravity effects. This is not the range of scales we shall be interested in, but rather macroscopic scales for which there is astrophysical evidence for BHs.} both in view of theoretical consistency, in particular dynamics, as well as in view of the current observational developments.

\section{Non-Kerrness: guiding principles and testing grounds}
\label{sec:2}

Non-Kerr models for astrophysical BH candidates must obey theoretical reasonability criteria. Here is a possible minimal list~\cite{Degollado:2018ypf}: \\ 1) to appear in well motivated and consistent physical theories. In the case of Kerr, this is vacuum General Relativity (GR), which, albeit an incomplete theory, due $e.g.$ to singularities, within its regime of validity fulfills this criterion; \\ 
2) to have a dynamical formation mechanism, which for Kerr is gravitational collapse~\cite{Nakamura:1987zz}, together with accretion and mergers; \\ 
3) to be sufficiently stable, meaning it can play a role in astrophysical or cosmological time scales. For Kerr, mode stability has been established long ago~\cite{Whiting:1988vc}.  \\
The two last criteria establish \textit{dynamical robustness}. This is one of the unifying principles of the models discussed below: there should be a route for forming them and be sufficiently stable against the unavoidable perturbations in any realistic environment. This is a restrictive criterion. Some models of alternative BHs or horizonless compact objects, $e.g.$ wormholes, have no established formation mechanism.

Being a good theoretical model in the sense of the previous paragraph is not, however, enough to describe Nature. The model must give rise to all the correct phenomenology attributed to astrophysical BHs, both in the electromagnetic and GWs channels. At the moment, there is not (yet?) clear tension between observations and the Kerr model. But the limitations of GR ($e.g.$ the unavoidable singularities behind BH horizons~\cite{Penrose:1964wq}) and of the standard model (SM) of particle physics (which fails to explain, say, dark matter) leverage us to consider non-Kerr models and how much we can distinguish them from the paradigm, with state of the art observables. Moreover, we should bear in mind that, at the moment, we are not testing simultaneously ($i.e.$ for the same mass range) all these observables. Thus, degeneracy in any single observable (even if not in all), for some model, may be of interest.

The first class of state of the art strong gravity observables, to test against,  comes from GWs. Amongst the many detections, a most interesting (and intriguing) one is  GW190521~\cite{LIGOScientific:2020iuh}. Key novel aspects of this event include: $i)$ the very massive progenitors, of about 85 and 66 $M_\odot$, meaning that, within the uncertainties, at least one is in the so called "pair instability supernova gap", a gap in the mass spectrum starting somewhere between 45 and 55 $M_\odot$ wherein BHs cannot form from the collapse of massive stars according to standard stellar evolution~\cite{LIGOScientific:2020ufj}. So, how did the progenitors of this event come to be? $ii)$ it is a low frequency event, implying there is no inspiral in the observed signal, leaving room to speculate about the true nature of the event. Thus, GW190521 (and similar events) promises to be a fertile ground for theoretical modelling, in particular for testing the Kerr hypothesis.

A second remarkable observable is the first image of a BH - M87* -  resolving horizon scale structure~\cite{EventHorizonTelescope:2019dse}. Since this observation is probing the strong gravity region of a few Schwarzschild radii, even though it has only been done for a single BH and even though there may be a non negligible impact of the astrophysical environment, which is not fully under control, it is worth understanding how it can be used to confirm the Kerr hypothesis or constrain deviations thereof, by comparing the shadow~\cite{Falcke:1999pj,Cunha:2018acu} and emission ring in the M87* image with non-Kerr models.

\section{Non-Kerrness: two families of examples}
\label{sec:3}

The Carter-Robinson uniqueness theorem~\cite{Carter:1971zc,Robinson:1975bv,Chrusciel:2012jk} establishes that physically reasonable BHs in vacuum GR fall into the family~\eqref{kerrm}. Hence they are rather featureless; they have \textit{no-hair}~\cite{Ruffini:1971bza}. Thus, to find non-Kerr BHs, one must either include matter in GR or consider modified gravity. This is necessary but not sufficient. For instance, including minimally coupled matter fields, with standard kinetic terms and obeying (say) the dominant energy condition is rather restrictive; it prevents non-Kerr BHs in many models. This conclusion has been established by model-specific no-hair theorems~\cite{Herdeiro:2015waa,Volkov:2016ehx}.  Some of these theorems are  powerful since they do not require much, a paradigmatic example being Bekenstein's theorem ruling out BH ``hair" of a real scalar field minimally coupled to Einstein’s gravity, possibly with a mass term or even some classes of self interactions~\cite{Bekenstein:1972ny}. Still, theorems have assumptions, and dropping some of them one may (sometimes) find reasonable scenarios where new BHs emerge.

Here, we shall be interested both in examples in GR with minimally coupled scalar (or massive vector) fields, where the usual theorems are circumvented in a subtle way, by  a property called ``symmetry non-inheritance", and in models beyond GR where the theorems are circumvented by the use of non-minimal couplings. In either case, the theoretical foundation will invoke new physics which introduces a new scale. Moreover, in both cases, the non-Kerr BHs co-exist with Kerr as solutions of the model. At some scales, however, the former emerge dynamically from the latter. In other words, the Kerr solution becomes \textit{unstable} due to some new physics and a new preferred state emerges. 

The first family is in GR but with matter beyond the SM, in fact ultralight bosonic particles, that have been proposed as dark matter candidates~\cite{Hui:2016ltb,Suarez:2013iw,Freitas:2021cfi}. The two basic members of this family are described by the action ($G=1=c$)
\begin{equation}
    S^{(s)}=\int  d^4x \sqrt{-g}\left[  \frac{R}{16\pi}+\mathcal{L}^{(s)}
 \right] ,
 \label{model1}
\end{equation}
where $\mathcal{L}^{(0)}=
   - \bar{\Phi}_{, \, \alpha} \Phi^{, \, \alpha}   - \mu^2 \bar{\Phi}\Phi$ and $\mathcal{L}^{(1)}=
   -\bar{F}_{\alpha \beta}F^{\alpha\beta}/4-\mu^2\bar{A}_\alpha A^\alpha/2$ for the scalar ($s=0$) and vector ($s=1$) cases, respectively. $\Phi$ and $A_\mu$ are a complex scalar and vector field, $F=dA$ and overbar denotes complex conjugate. The field mass introduces a new scale, $\mu$, which is an inverse length, and which, for the scenario herein, is taken to be \textit{ultralight}, with mass between  $10^{-10}$ to $10^{-20}$ eV. Kerr BHs with a horizon scale comparable to the Compton wavelength of this new particle, become (efficiently) unstable against a process called \textit{superradiance}~\cite{Brito:2015oca}, transferring part of their energy and angular momentum to a scalar cloud around the BH, which becomes  a "BH  with synchronised (scalar or Proca) hair”~\cite{East:2017ovw,Herdeiro:2017phl}, and different from Kerr. These BHs were first reported in~\cite{Herdeiro:2014goa} and~\cite{Herdeiro:2016tmi} for the scalar and Proca case, respectively.
   
   The second family is beyond GR, in modified gravity. A member of this family is the extended scalar-tensor Gauss-Bonnet (eSTGB) model, described by the action
   \begin{equation}
       S=
\frac{1}{16\pi}\int d^4x \sqrt{-g} \left[  R - 2\partial_\mu \phi \partial^\mu \phi
 +  \lambda^2f(\phi)(R^{\mu\nu\alpha\beta}R_{\mu\nu\alpha\beta}-4R_{\mu\nu}R^{\mu\nu} +R^2)  \right], 
 \label{estgb}
   \end{equation}
   where $\phi$ is a real scalar field and $f(\phi)$ an yet unspecified coupling function.    The Gauss-Bonnet (GB) coupling introduces a new scale $\lambda$, which has units of length. Then, BHs with a horizon scale comparable to this new scale, can undergo a strong gravity phase transition, \textit{spontaneous scalarisation}~\cite{Doneva:2017bvd,Silva:2017uqg}.\footnote{Spontaneous scalarisation was first proposed for neutron stars, in a different model~\cite{Damour:1993hw}.}  This leads, dynamically, to new types of BHs, dubbed "scalarised BHs". 

Model~\eqref{model1} also accommodates horizonless compact objects known as "bosonic  stars", scalar~\cite{Kaup:1968zz,Ruffini:1969qy,Schunck:2003kk} or vector~\cite{Brito:2015pxa,Herdeiro:2017fhv,Herdeiro:2019mbz}. The latter are also known as "Proca stars".   Some of these solutions are dynamically robust~\cite{Liebling:2012fv,Sanchis-Gual:2019ljs} and have a formation mechanism, via a process called "gravitational cooling"~\cite{Seidel:1993zk,Guzman:2006yc,DiGiovanni:2018bvo,Sanchis-Gual:2019ljs}. When the new scale, $\mu$ is in the aforementioned range, between  $10^{-10}$ to $10^{-20}$ eV, the maximal mass of these bosonic stars is in the astrophysical range of $1-10^{10}$ $M_\odot$, and these objects mimic the mass of astrophysical BHs. Scalar and vector bosonic stars  have some interesting differences which will be emphasised in the examples below concerning the \textit{(BH) imitation game}.

\section{Non-Kerr BHs: the example of synchronisation}
\label{sec:4}

\subsection{Dynamical considerations}

Dynamical synchronisation occurs in many systems, both in biology, $e.g.$ communities of fireflies or crickets, and in physics, $e.g.$ sets of metronomes or pendulums~\cite{strogaatz}. In these systems, individual cycles converge dynamically to the same phase due to appropriate interactions, yielding a configuration that would otherwise look fine-tuned. 

In Newtonian gravitational dynamics, synchronisation occurs in binary systems of extended objects, such as planets or stars~\cite{Hut}. Tidal effects tend to synchronise and lock orbital and rotational periods. This effect led the moon to always show the same face towards the Earth, and the Earth is (very slowly) tending to show the same face towards the moon. This is a ubiquitous behaviour observed in all planets-moons of the solar system.

In the context of relativistic gravity, the aforementioned BHs with synchronised bosonic hair can be interpreted as synchronised configurations (hence the name). The synchronisation condition reads $\Omega_H=\omega/m$~\cite{Hod:2012px,Dias:2011at,Herdeiro:2014goa}, where $\Omega_H$ is the horizon angular velocity, and $\omega/m$ is the phase angular velocity of the bosonic field, which has a dependence $\Psi\sim e^{-i(\omega t-m\varphi)}$, where $\Psi$ represents either $\Phi$ or $A_\mu$; $t,\varphi$ are the time and azimuthal coordinates of the stationary and axisymmetric spacetime; $\omega$ is the frequency of the bosonic field’s harmonic time dependence  and $m$ is an integer azimuthal harmonic index.
This condition is stating that the phase of the field is co-rotating in synchrony with the horizon. Are these synchronous hairy BHs attained dynamically, as in the case of the fireflies or pendulums?

There is indeed one dynamical channel that leads to synchronisation: the process of superradiance. Fully non-linear numerical simulations have been successfully performed when the bosonic field is the Proca one~\cite{East:2017ovw}. They have shown that the process of superradiant rotational energy extraction spins down the BH, saturates and a new equilibrium state is attained, precisely when the BH and the dominant superradiant mode obey  $\Omega_H=\omega/m$. These simulations obtained a maximum of $\sim$ 9\% of energy transfer from the BH into the bosonic cloud. This is close to the maximum expected in the evolution of the dominant superradiant mode, which is $\sim 10\%$, also for the scalar case~\cite{Herdeiro:2021znw}. Using the data from these simulations the new equilibrium state was identified with a BH with synchronised Proca hair~\cite{Herdeiro:2017phl}.

We thus have a process creating a new sort of BH, which takes some time scale. But the time scale depends crucially on a resonance between the Compton wavelength  of the fundamental boson $1/\mu$ and the Schwarzschild radius of the BH, $\sim M$. Maximal efficiency occurs when  they are  similar, $M\mu\sim 1$~\cite{Dolan:2012yt}.  Otherwise the time  scale quickly grows and becomes larger than the Cosmological time sufficiently far from the sweet spot. This therefore selects a mass scale of BHs: depending on  the sort of mass  of the fundamental boson, BHs in a certain (narrow) mass range become  hairy, but outside this range they (effectively) do not~\cite{Degollado:2018ypf}. The punch line is that, under the assumption a single ultralight boson with some mass $\mu$ exists, non-Kerrness would manifest itself only for BHs in some narrow mass range around $1/\mu$. Observational evidence for such non-Kerrness would therefore identify $\mu$.

There may be other formation channels for these synchronised BHs, in particular,   from mergers of bosonic stars, both scalar and vector~\cite{Sanchis-Gual:2020mzb}. There is a key difference in this latter channel: superradiance forms a synchronised configuration by spinning down the BH.  In this new channel, mergers of bosonic stars lead to a synchronised system by spinning up the BH that results from the merger, and which accretes part of the field remnant. In the latter case, however, fine tuning seems necessary~\cite{Sanchis-Gual:2020mzb}.

\subsection{Comparison with observations}

Can we constrain BHs with synchronised hair with current observations? Concerning GWs, both the perturbation theory (for the ringdown) and fully non-linear dynamical evolutions (for the inspiral and merger) remain essentially unexplored (but see ~\cite{Collodel:2021jwi}). Concerning shadows, on the other hand, more progress has been achieved, $e.g.$~\cite{Cunha:2015yba,Cunha:2019ikd}.  

\begin{figure}[h!]
\includegraphics[scale=.059]{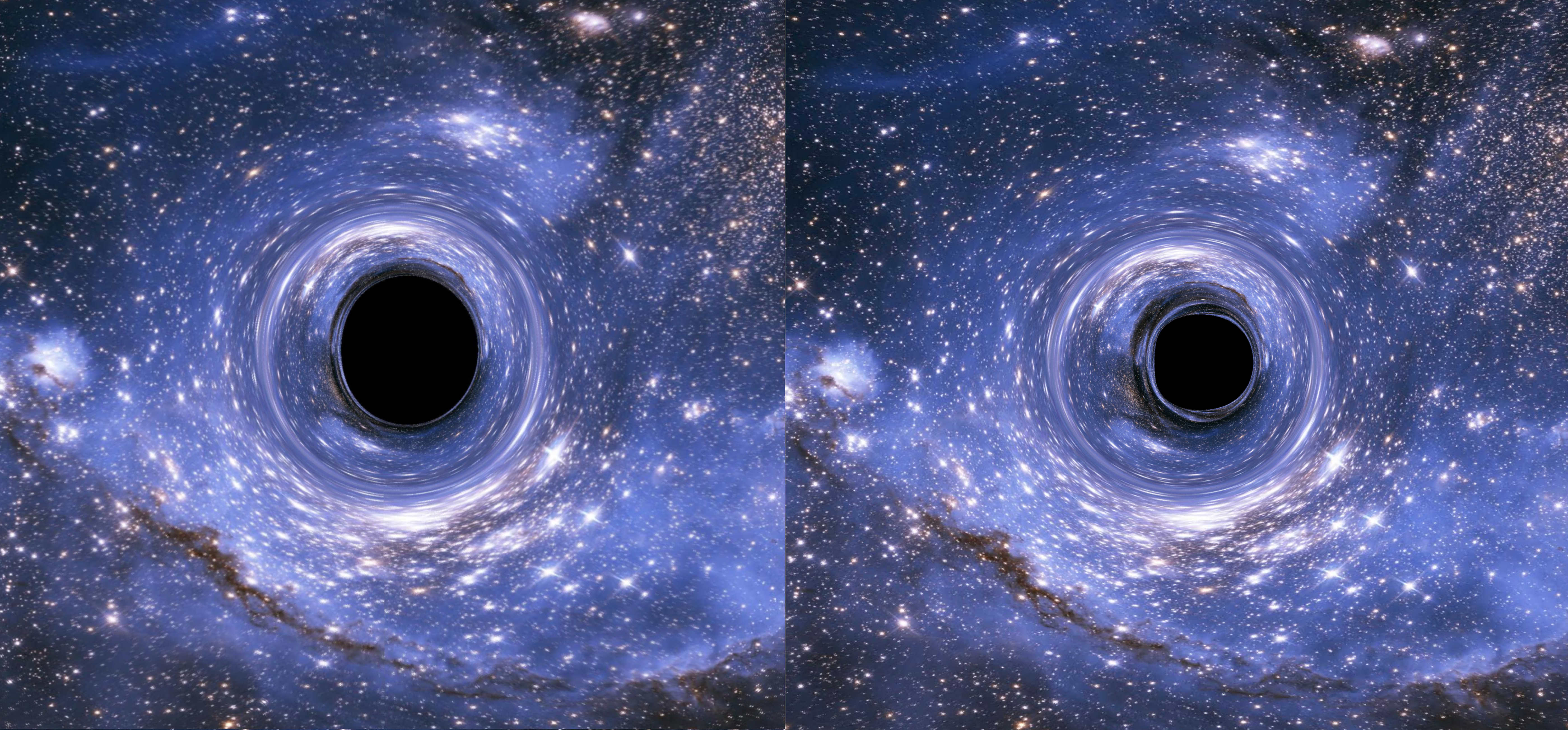}
%
%
\caption{Shadow and lensing of a Kerr BH (left) and a BH with synchronised scalar hair (right) with the same ADM mass and angular momentum, using a background image of Infant Stars in the Small Magellanic cloud from the Hubble space telescope. Adapted from~\cite{Cunha:2015yba}.}
\label{fig:1}       
\end{figure}

This family of BHs with synchronised hair interpolates between vacuum Kerr BHs and bosonic stars. Consequently, close to the Kerr limit phenomenological differences are as small as desired (compared to Kerr) due to the continuity of the solutions; but sufficiently far away they are large and can become huge in the solution space region close to bosonic stars.  Fig.~\ref{fig:1} shows lensing images obtained by ray tracing for these BHs and comparing with Kerr~\cite{Cunha:2016bpi}, using an aesthetically appealing starry sky as the light source, rather than a realistic astrophysical environment.  The hairy BH shown (righ panel) has the same mass and angular momentum as the Kerr one (left panel), but  75\% of the  mass and 85\% of the angular momentum are stored in the scalar field (rather than the horizon). One can see that due to the transfer of part of the energy and spin of the BH to the  scalar “hair” the actual shadow is considerably smaller than that of a Kerr BH with the same total mass and spin and  comparable observation conditions. Actually it is about 25\% smaller in terms of the average radius, which seems to exclude this particular example of hairy BH as a good  model for M87* with current EHT observations. This illustrates how sufficiently large departures from Kerr can be falsified even with current observations. However, how are the differences in the dynamically viable region, assuming formation from superradiance?

In the dynamically most interesting region of the parameter space, where the hairy BHs can form from Kerr in an astrophysical time scale and are stable (against higher superradiant modes) for at least a Hubble time~\cite{Degollado:2018ypf}, the differences are generically small. Recall that only about $\sim$10\% of the Kerr BH mass can be transferred to the bosonic cloud. Yet…this is a non-negligible amount of mass for a 6 billion solar masses BH like M87*. Analysing in this region of the parameter space if the M87* image could then distinguish the hairy BH from Kerr, one concludes  that  all  dynamically viable BHs could  be mistaken for a Kerr BH  within the  current error bars~\cite{Cunha:2019ikd}. This conclusion seems quite  natural, since roughly, the error in the measurement of the emission ring diameter is about 10\%. 

The conclusion is the following. Consider that an ultralight (dark matter) particle with a mass of $\sim10^{-20}$ eV exists in Nature, so that $\mu M_{M87*}\sim 1$. Then, as M87* grew throughout its history to its current mass, superradiance became efficient around its current mass, and endowed it with synchronised hair, a process that took a few million years. The (current) hairy M87* BH is effectively stable against higher superradiant modes, for a Hubble time.  In this scenario,  M87* is not a Kerr BH, but rather a  BH with synchronised hair. However, the current Event Horizon Telescope precision would not be able to tell the difference. Observe, moreover, that all other BHs with a mass smaller than $\lesssim 10^9 M_\odot$ would not become hairy, as superradiance is not efficient. They would be Kerr BHs.

\section{Non-Kerr BHs: the example of scalarisation}
\label{sec:5}

We now consider a different scenario wherein non-Kerr BHs also arise dynamically, for some scales, but via a different instability of Kerr BHs: spontaneous scalarisation. 

In the landscape of modified gravity models there are question marks on the well-posedness and theoretical  consistency of many of them. A fairly well motivated class of models with higher curvature corrections is the class of eSTGB models~\eqref{estgb}. Here, the model is defined by the choice of the coupling between the GB and the scalar field, $f(\phi)$. Choosing a dilatonic coupling, $f(\phi)\sim e^{\beta \phi}$, which is motivated, say, by string theory or dimensional reduction, Schwarzschild/Kerr are not solutions; there are new BHs which are stable (in some regime) and which have a qualitatively new feature, namely  a minimal BH size~\cite{Kanti:1995vq,Kleihaus:2015aje}. This can  be interpreted as due to a repulsive effect, sourced by the GB term, that destabilises the horizon when the GB term  becomes dominant,  $i.e.$ for  sufficiently small BHs. 

Changing the scalar-curvature coupling into a more general $f(\phi)$, there are  various interesting cousin models. For example, one can consider shift-symmetric models, which are close to the previous  dilatonic model  but with this additional shift-symmetry. BHs in this model have been constructed~\cite{Sotiriou:2014pfa,Delgado:2020rev} and shown to exhibit dynamical formation~\cite{Benkel:2016kcq}. On the other hand, the condition $df/d\phi(\phi=0)=0$,  yields a class of models admitting \textit{both} vacuum GR and scalarised BHs~\cite{Doneva:2017bvd,Silva:2017uqg,Antoniou:2017acq}.  Moreover, depending on the sign of $d^2f/d\phi^2(\phi=0)$, the GR BHs may be unstable in some scales, $i.e.$ when $M/\lambda$ is smaller than some threshold, which suggests the non-GR BHs could emerge dynamically, via spontaneous scalarisation. Entropic considerations support this possibility in some models, $e.g.$~\cite{Doneva:2017bvd,Cunha:2019dwb}. 

In the case of these models, the fully dynamical process has not been established, but there are dynamical results showing the exponential growth and saturation of a self-interacting scalar field in the decoupling limit~\cite{Doneva:2021dqn}. Moreover, fully non-linear numerical evolutions could be performed of the spontaneous scalarisation of a Reissner-Nordstr\"om BH in a cousin model, showing the phenomenon occurs dynamically and leads to a perturbatively stable scalarised BH~\cite{Herdeiro:2018wub}.

The scalarised Kerr BHs that would be the endpoint of the scalarisation process were first constructed in~\cite{Cunha:2019dwb} (see also~\cite{Collodel:2019kkx}), in model~\eqref{estgb} with  $f(\phi)=(1-e^{-\beta \phi^2)}/(2\beta)$. Choosing this coupling is illustrative. Some properties (but not all) of  the scalarised solutions are universal for all couplings allowing scalarisation, $e.g.$ the threshold between Kerr and scalarised BHs. All solutions (as in the case of the BHs with synchronised hair) are numerical. Consider first static, spherical BHs, taking $\lambda$ as fixed (new)  scale. For sufficiently large BHs  as compared to $\lambda$ (greater than $\sim$ 0.6) there are  only  Schwarzschild BHs. No scalarised BHs exist.  Decreasing the  mass of the BH, Schwarzschild BHs becomes unstable against  scalarisation and the  scalarised BHs appear. These  are stable, at least against radial perturbations up to some smaller  size, where the scalarised BHs cease to be stable~\cite{Blazquez-Salcedo:2018jnn}. So, there is a window of sizes where scalarised BHs are stable and dynamically preferred to Schwarzschild. 

\begin{figure}[h!]
\centering
\includegraphics[scale=.34]{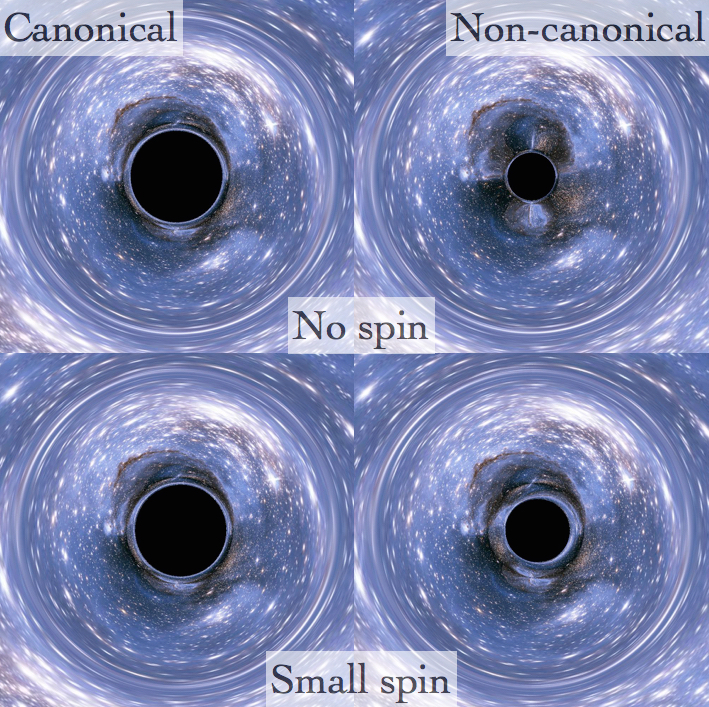}
%
%
\caption{Shadow and lensing of scalarised BHs (right panels) and Schwarzschild/Kerr BHs (right panels) with the same ADM mass and angular momentum, using the same background image as in Fig.~\ref{fig:1}. Adapted from~\cite{Cunha:2019dwb}.}
\label{fig:2}       
\end{figure}

The effect of the angular momentum is interesting:  spin  suppresses  the effects  of scalarisation~\cite{Cunha:2019dwb}. This spin suppression of the non-Kerrness can be seen using the diagnosis of the associated shadows - Fig~\ref{fig:2}. Using the same background image as before, one sees that for non-spinning  BHs (top panels), the distinction is visible with a naked eye:    the Schwarzschild BH shadow (left panel) and that of a scalarised BH with the same mass and under similar observation conditions (right panel) are clearly different. The latter is in the perturbatively stable region. But as the spin parameter increases (bottom panels) the difference is suppressed  and even for still fairly low spins, say $j\sim 0.5$, they become negligible.

In ~\cite{Cunha:2019dwb} a comparison with the Event Horizon Telescope data was performed. Taking into account that very little is known about the spin of M87*, however, this comparison is not very informative. Even in the  most optimistic  scenario (for this model) where the spin is low, one can only put a rather weak constraint on the new scale that the model introduces.

Let us close this discussion on spontaneous scalarisation with a different class of models that introduces a new twist. In the models we have just discussed, the Kretschmann scalar of the vacuum BH is providing the instability, endowing scalar perturbations with a tachyonic mass. This Kretschmann scalar is positive for Schwarzschild. For Kerr it starts to become negative (around the poles) for the dimensionless spin, $j>0.5$. Thus, if we consider models with the opposite coupling sign, only sufficiently fast spinning BHs will scalarise. This has been called "spin induced scalarisation"~\cite{Dima:2020yac}.

In~\cite{Herdeiro:2020wei} (see also~\cite{Berti:2020kgk})  the domain of existence of the corresponding BHs was explored, for the same illustrative coupling as before. Whereas in the model before, spin quenched the Kerr deviations, now spin enhances the Kerr deviations;  in fact it is mandatory. So,  this illustrates how  there are models in which only some BHs, either with small or with large spin can differ from Kerr, which moreover only occurs for some mass scales. In the case of spin induced scalarisation a detailed phenomenological study of the solutions, namely of the shadows, has not yet been reported.

\section{The imitation game: non-BHs mimicking BH observables}
\label{sec:6}

Let us now discuss “the imitation game”, or how non-BHs can mimic some (even if not all) BH observables. 

There are several motivations to consider BH mimickers,  $e.g.$, the singularity problem of BHs. This has led to many models of horizonless compact objects that could behave as BH imitators.  It has been pointed out, however, that an imitator needs to have a \textit{light ring} (LR) to mimic the BH ringdown~\cite{Cardoso:2016rao}. LRs also play a key role in determining the edge of the shadow~\cite{Cunha:2017eoe}.  Thus, it seems that for compact object to mimic the most strong field current observations they need to possess LRs (and are then called "ultra-compact").   

There may be, however, a generic possible issue with horizonless ultracompact objects.  A theorem established in~\cite{Cunha:2017qtt} states that for generic equilibrium ultracompact objects, resulting from a smooth, incomplete gravitational collapse, (thus, for which there is plausible formation mechanism) LRs come in pairs and one is stable. A stable LR has been suggested to trigger a (non-linear) spacetime instability~\cite{Keir:2014oka}, as massless perturbations can pile up in its vicinity. Not much is known about this instability or its timescale, but it raises a shadow of doubt about the dynamical robustness of horizonless, ultracompact objects.

Let us therefore ask the question if real data, like a true GW event or the Event Horizon Telescope M87* observation could be imitated by a compact object mimicker that does not have LRs. Interestingly, as the following case study examples indicate the answer is yes in both cases.

\subsection{Mimicking a GW event}
\label{sec:61}

Let us first address the imitation of a GW event. In GR coupled to ultralight bosonic fields, there has been, for many years, developments in evolving scalar and vector boson stars dynamically, using numerical relativity techniques~\cite{Liebling:2012fv}. In 2019 an unexpected difference between spinning scalar and vector bosonic stars was found: the most fundamental spinning scalar boson stars develop a non-axisymmetric instability \cite{Sanchis-Gual:2019ljs}. When  this instability kicks in, these stars collapse into a BH. On the other hand, this instability is absent in the cousin Proca model. Indeed, even if one perturbs considerably a spinning Proca star, no instability is seen.  So, spinning Proca stars, without self-interactions, are dynamically robust, unlike their scalar cousins.

Given the dynamically robustness of spinning Proca Stars, recently we considered simulations of mergers of spinning Proca  stars and compared  them with real data, in particular with the intriguing event GW190521 commented on earlier. We have found, through a Bayesian analysis, that a collision of two spinning Proca stars actually fits slightly better the data than the vanilla binary BH model that was used by the LIGO-Virgo collaboration~\cite{Bustillo:2020syj}.

It follows that if, even just as a proof of concept, one takes seriously the Proca model, we  can use the data  to infer the mass of the ultralight bosonic (in this case vector) particle.  We have obtained a mass of around $8\times10^{-13}$ eV. More similar events are needed to  confirm this  possibility.  In the most likely case scenario this  is just a proof of concept showing 1) how there can be degeneracy in real data between two  very different models and 2) how one could extract physical information about a new  fundamental, dark matter particle from  GW data (in this case the mass of the  boson). Note that the colliding Proca stars are compact but not ultracompact, so they do not have LRs. Of  course, in a much more exciting possibility, a potential confirmation of the bosonic star scenario would be a first hint for the long sought dark matter nature. 

Under this rationale, one may ask why could this event be such a Proca star collision and not the other events detected so far? The point is again mass selection. The mass of the ultralight bosonic particle determines the maximal mass of the corresponding stars. For the above quoted ultralight boson, this mass turns out to be about $\sim$173 $M_\odot$. All other events correspond to a smaller final mass. Since we have observed BH formation, due to the ringdown, then these cannot be Proca star collisions. For the Proca star collisions to form a BH they have to overshoot this limit, which requires events as massive as GW190521.

\subsection{Mimicking a BH shadow without LRs}
\label{sec:61}

Let us now address the imitation of a BH shadow by a mimicker without LRs.

In a generic stationary and axi-symmetric BH spacetime, one can associate the edge of the BH shadow to a set of photon bound orbits, which we refer to as "fundamental photon orbits"~\cite{Cunha:2017eoe}. In the Schwarzschild case, all of  these are planar LRs. In the Kerr case  they are called "spherical photon orbits",  since  they have a constant  radial Boyer-Lindquist coordinate. 

In a real astrophysical environment, however, an effective shadow seen may depend on the details of the light source. This was nicely illustrated in~\cite{Olivares:2018abq}, where GR magnetohydrodynamic (GRMHD) simulations were performed on the background of BHs and of some models of static scalar boson stars. None of these stars have LRs (or a horizon); yet, some could produce an effective shadow, where others did not. In all boson stars models considered, they admitted stable timelike circular orbits until their very centre;  but in the case where an effective shadow was seen the angular velocity of the timelike circular orbits attains a  maximum at some non-zero areal radius $R_\Omega$. This new scale is observed to  determine  the inner edge of the accretion disk in the simulations, under some assumptions, including that the loss of angular momentum of the orbiting matter is driven  by  the  magneto-rotational  instability and  that the radiation relevant for the BH shadow observations is mostly due to synchroton emission.  There are, however, two caveats for the models in~\cite{Olivares:2018abq} that produce an effective shadow, to be seriously considered as imitators. First, the imitation is not perfect, since the efective shadow is considerably smaller than that of a comparable Scwharzschild  BH with the same mass. Secondly, the boson stars producing an effective shadow are perturbatively unstable.

It turns out, as discussed in~\cite{Herdeiro:2021lwl}, that this imitation game works better for Proca stars (rather than scalar boson stars).  Within the stable branch of spherical, fundamental Proca stars, there are solutions that display the necessary new scale, that is, for which the timelike circular orbit with the maximal angular velocity is at some radius $R_\Omega\neq 0$. One can even choose 
a particular solution for which this new scale equals the location of the Innermost Stable Circular Orbit (ISCO) of a Schwarzschild BH with the same mass.  So, there is a dynamically robust solution that, under some accretion models yields an accretion disk morphology mimicking that of a Schwarzschild BH of the same mass. 

\begin{figure}[h!]
\centering
\includegraphics[scale=0.080]{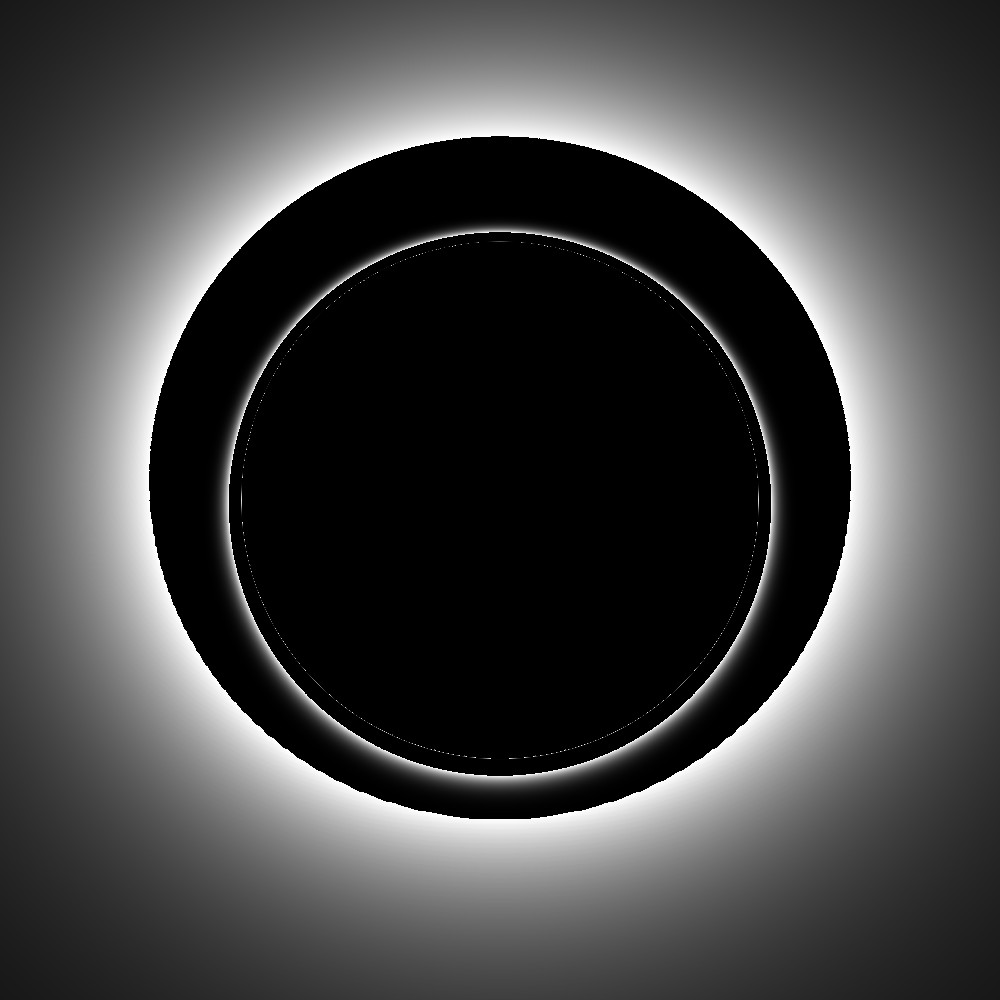}
\includegraphics[scale=0.080]{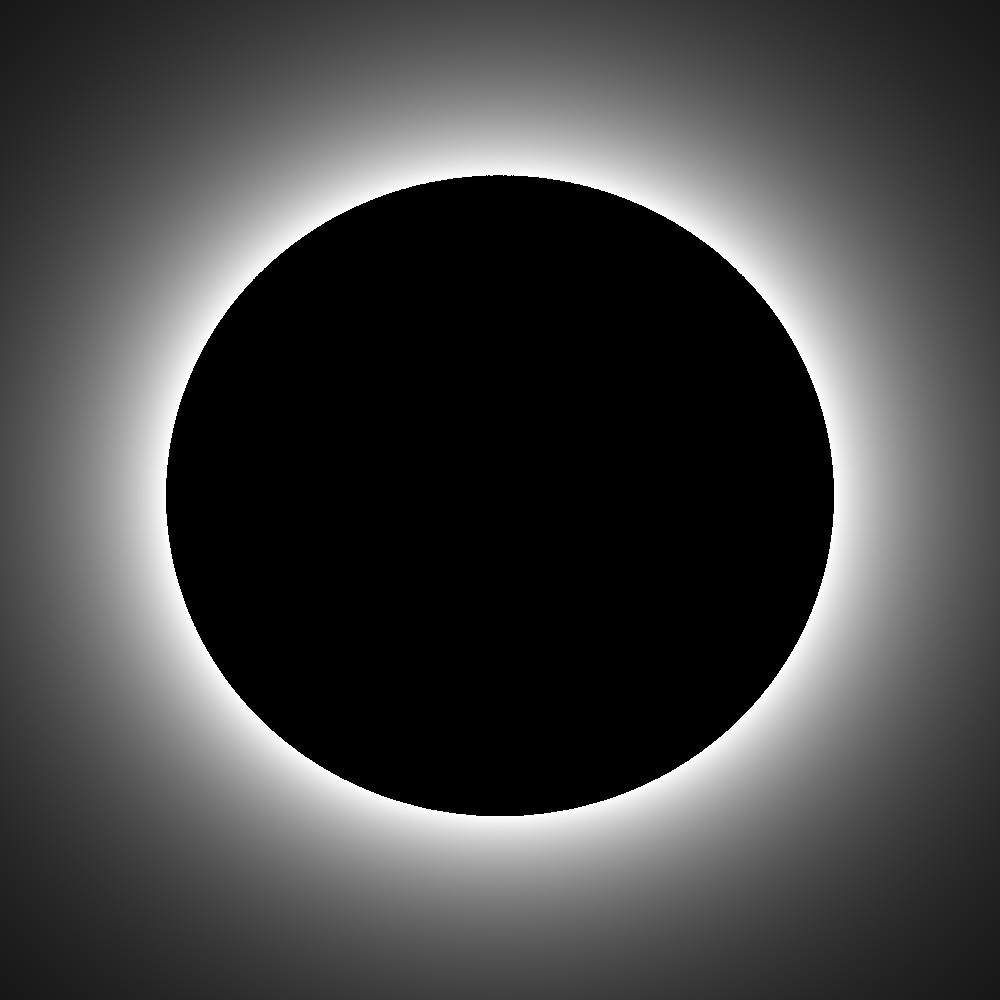} \ \ 
\includegraphics[scale=0.08]{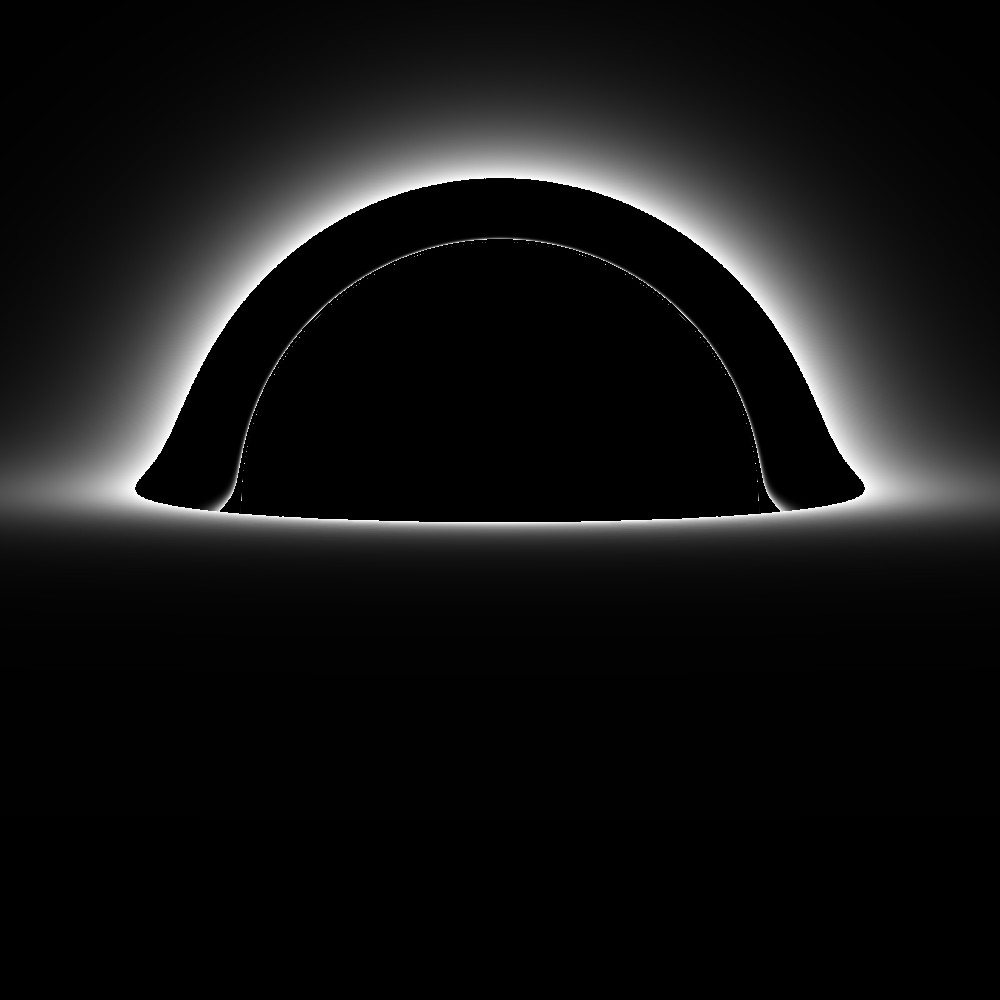}
\includegraphics[scale=0.08]{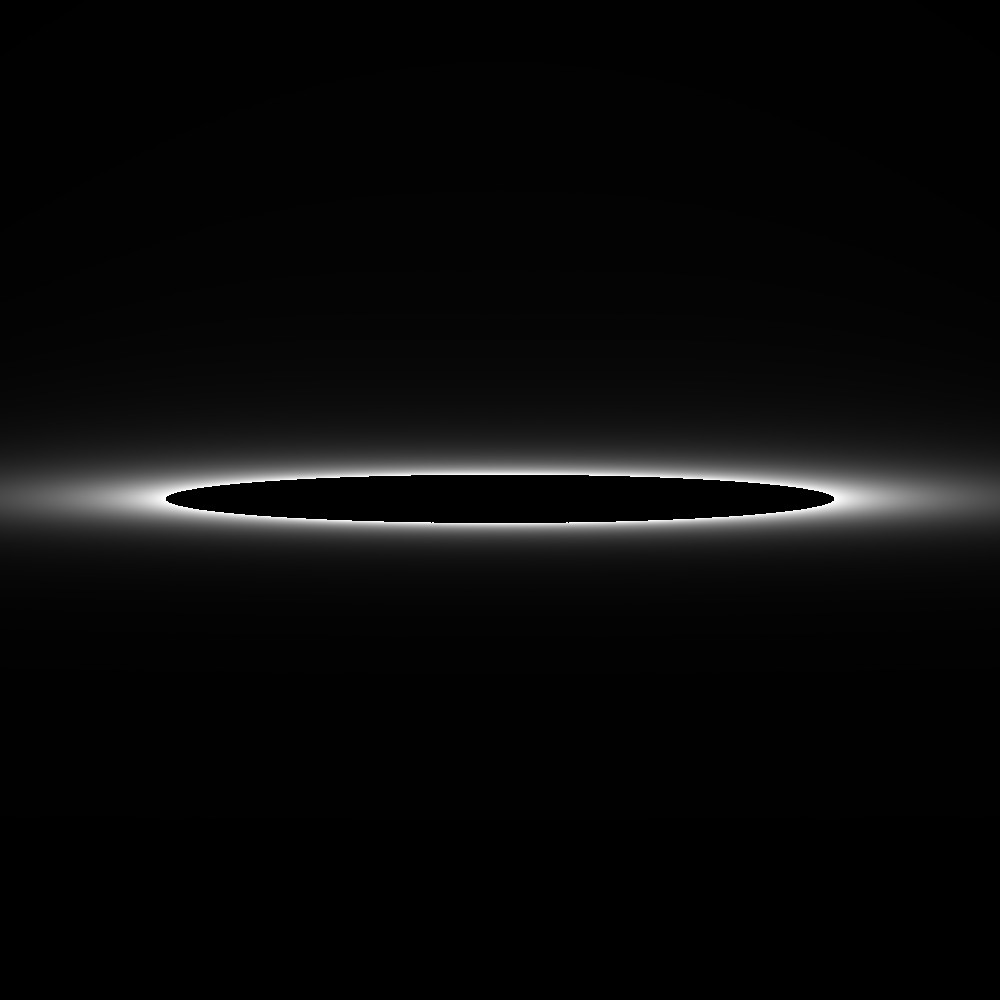}
\includegraphics[scale=0.08]{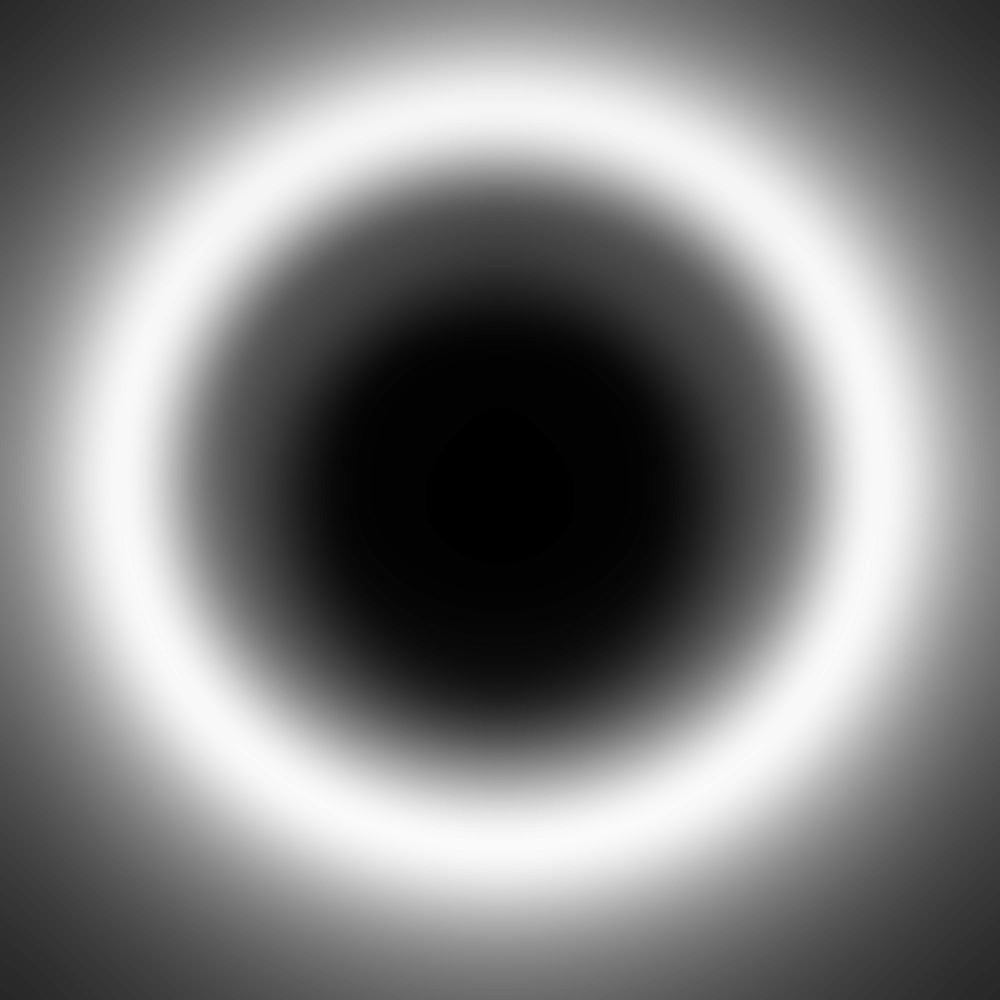}
\includegraphics[scale=0.08]{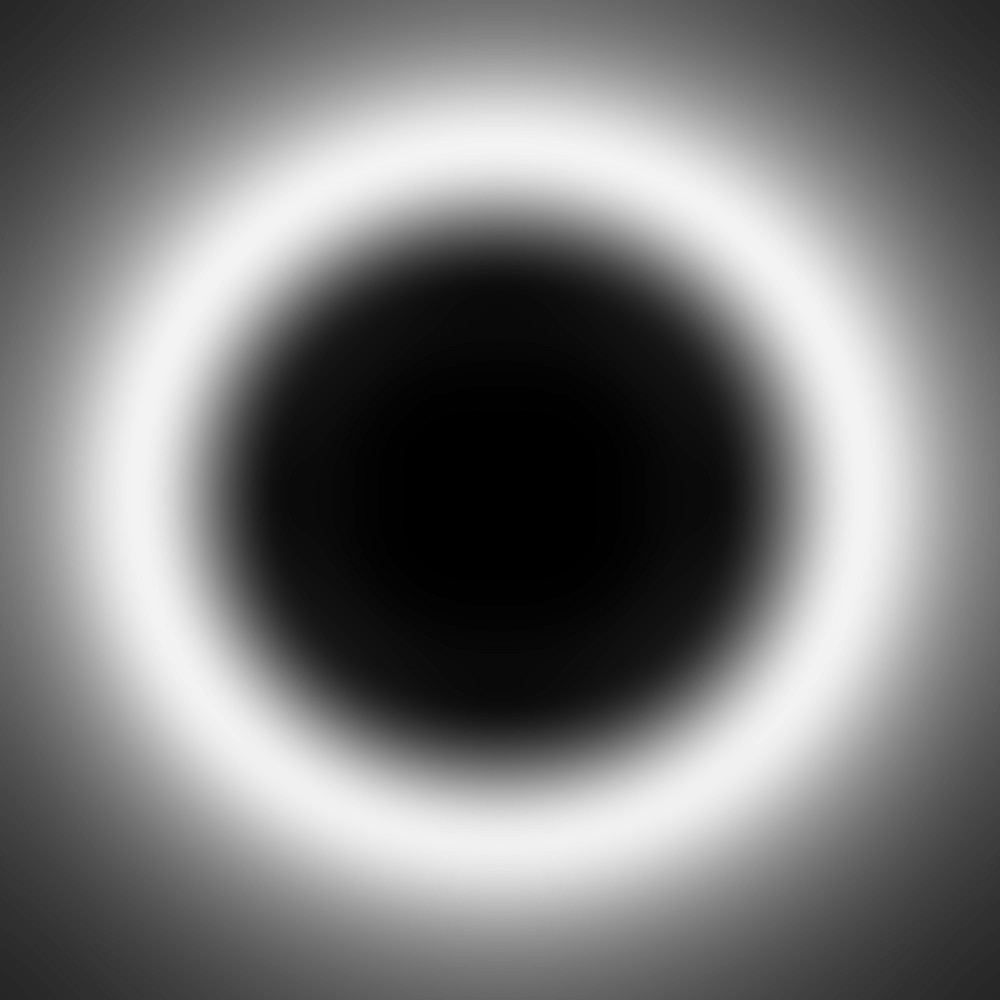} \ \ 
\includegraphics[scale=0.08]{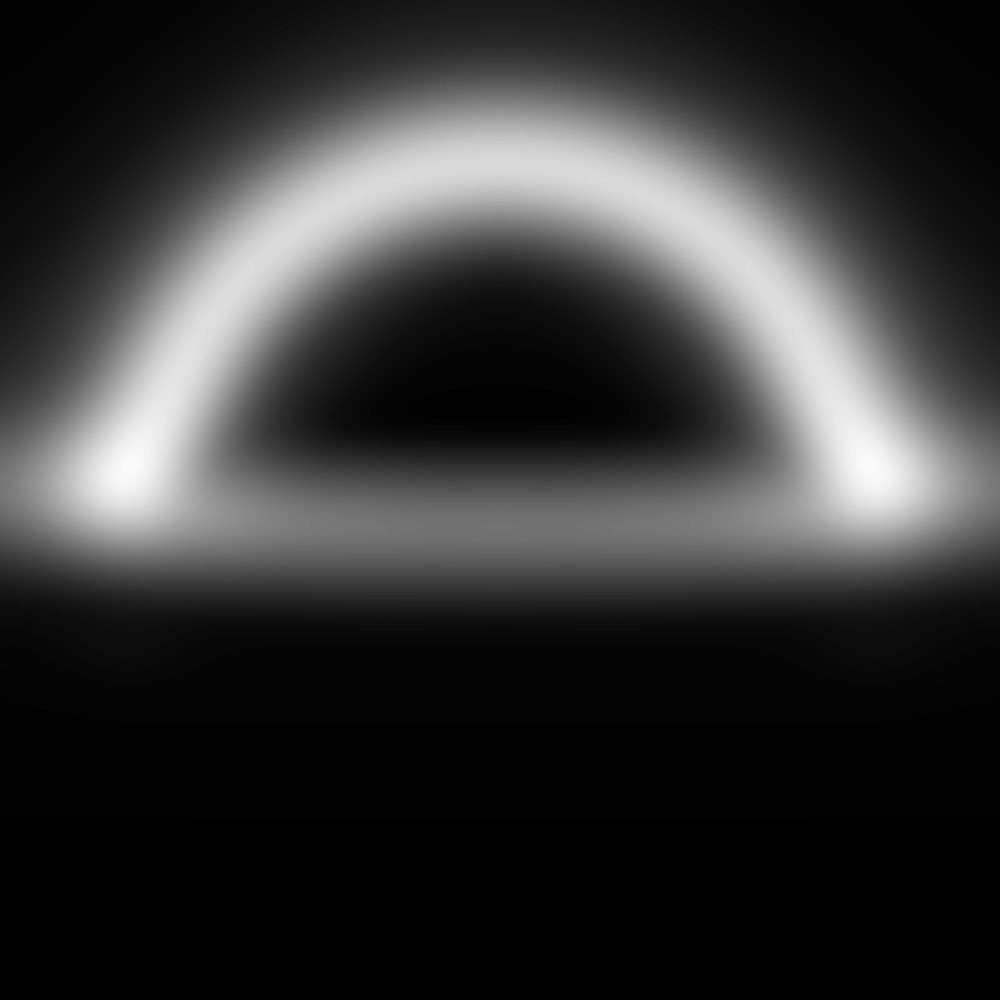}
\includegraphics[scale=0.08]{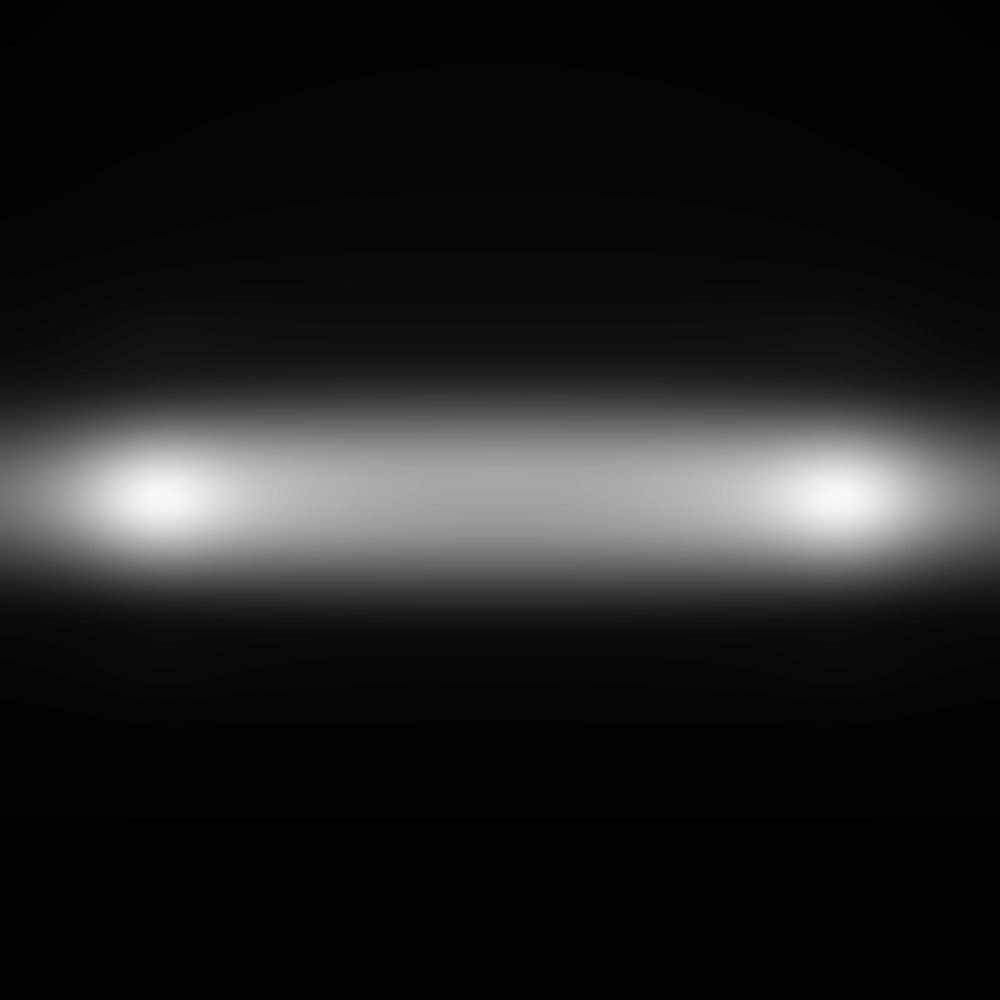}     
%
%
\caption{Shadow and lensing of the Proca star discussed in the text and of a comparable Schwarzschild BH, both illuminated by a thin, equatorial accretion disk, at two different observation angles. The bottom images are blurred to mimic current observational limitations.  Adapted from~\cite{Herdeiro:2021lwl}.}
\label{fig:3}       
\end{figure}

To check the potential shadow degeneracy, in~\cite{Herdeiro:2021lwl} ray tracing images were produced considering a simplified astrophysical setup, wherein the only radiation source is an opaque and thin accretion disk located on the equatorial plane around the central compact object. The disk has an inner edge with an areal radius $R_{\Omega}=6M$ in both spacetimes. The most interesting case for degeneracy occurs for an observer close to the poles (to match the estimated angle at which M87*, was observed from Earth) - Fig.~\ref{fig:3}. For this angle, the images of the  Schwarzschild  (top left panel) and Proca star (top middle left panel) look similar, although some finer additional lensing features are still visible in the Schwarzschild case. But these subtle differences are washed away if one applies a Gaussian blurring filter to the images, to mimic the current observations limited angular resolution, of the order of the compact object itself. The images so obtained are shown below the corresponding unblurred image, and are essentially indistinguishable. 

On the other hand, a near equatorial observation leads to fairly different images. The Schwarzschild one (top middle right panel) resembles the BH shape displayed in the  Hollywood movie “Interstellar”, whereas the Proca star simply looks like a flat accretion disk with a hole in it, as seen from the side (top right panel). This is because the gravitational potential well of the Proca star is shallow and so the bending of light it produces is weak. Consequently, the accretion disk has an almost flat spacetime appearance. Applying the same blurring as before, even with limited resolution, the two objects could be distinguished.  Let us stress that full GRMHD analysis and ray-tracing in the background of this Proca star are required to fully settle the question: to what degree can it imitate a BH observation? The case built herein, nonetheless, clearly confirms a potential degeneracy, but only under some observation conditions. And these Proca stars do not have LRs and are dynamically stable.

\section{Concluding remarks}
\label{sec:7}

Let us close with two following final remarks.  Firstly, all these models we have discussed have caveats; but they illustrate theoretical possibilities of dynamically robust non-Kerr BHs, or BH imitators, that could manifest themselves only at some specific scales of  mass and spin. Secondly, that producing detailed phenomenology will constrain the model and the corresponding (admittedly exotic) physics or, in the best case scenario, provide a smoking gun to this new physics.

\begin{acknowledgement}
I thank all my collaborators that were fundamental for the work presented here. Special thanks, by virtue of our long and fruitful collaboration go to P. Cunha, E. Radu and N. Sanchis-Gual. 
This work is supported by the Center for Research and Development in Mathematics and Applications (CIDMA) through the Portuguese Foundation for Science and Technology (FCT - Funda\c{c}\~ao para a Ci\^encia e a Tecnologia), references UIDB/04106/2020, UIDP/04106/2020 and the projects PTDC/FIS-OUT/28407/2017, CERN/FIS-PAR/0027/2019,  PTDC/FIS-AST/3041/2020 and CERN/FIS-PAR/0024/2021. This work has further been supported by  the  European  Union's  Horizon  2020  research  and  innovation  (RISE) programme H2020-MSCA-RISE-2017 Grant No.~FunFiCO-777740.

\end{acknowledgement}

\bibliographystyle{unsrt}
\bibliography{refs} 

\begin{thebibliography}{10}

\bibitem{Kerr:1963ud}
Roy~P. Kerr.
\newblock {Gravitational field of a spinning mass as an example of
  algebraically special metrics}.
\newblock {\em Phys. Rev. Lett.}, 11:237--238, 1963.

\bibitem{Boyer:1966qh}
Robert~H. Boyer and Richard~W. Lindquist.
\newblock {Maximal analytic extension of the Kerr metric}.
\newblock {\em J. Math. Phys.}, 8:265, 1967.

\bibitem{Narayan:2013gca}
Ramesh Narayan and Jeffrey~E. McClintock.
\newblock {Observational Evidence for Black Holes}.
\newblock 12 2013.

\bibitem{LIGOScientific:2018mvr}
B.~P. Abbott et~al.
\newblock {GWTC-1: A Gravitational-Wave Transient Catalog of Compact Binary
  Mergers Observed by LIGO and Virgo during the First and Second Observing
  Runs}.
\newblock {\em Phys. Rev. X}, 9(3):031040, 2019.

\bibitem{LIGOScientific:2020ibl}
R.~Abbott et~al.
\newblock {GWTC-2: Compact Binary Coalescences Observed by LIGO and Virgo
  During the First Half of the Third Observing Run}.
\newblock {\em Phys. Rev. X}, 11:021053, 2021.

\bibitem{LIGOScientific:2021djp}
R.~Abbott et~al.
\newblock {GWTC-3: Compact Binary Coalescences Observed by LIGO and Virgo
  During the Second Part of the Third Observing Run}.
\newblock 11 2021.

\bibitem{EventHorizonTelescope:2019dse}
Kazunori Akiyama et~al.
\newblock {First M87 Event Horizon Telescope Results. I. The Shadow of the
  Supermassive Black Hole}.
\newblock {\em Astrophys. J. Lett.}, 875:L1, 2019.

\bibitem{Gillessen:2008qv}
S.~Gillessen, F.~Eisenhauer, S.~Trippe, T.~Alexander, R.~Genzel, F.~Martins,
  and T.~Ott.
\newblock {Monitoring stellar orbits around the Massive Black Hole in the
  Galactic Center}.
\newblock {\em Astrophys. J.}, 692:1075--1109, 2009.

\bibitem{Ghez:2008ms}
A.~M. Ghez et~al.
\newblock {Measuring Distance and Properties of the Milky Way's Central
  Supermassive Black Hole with Stellar Orbits}.
\newblock {\em Astrophys. J.}, 689:1044--1062, 2008.

\bibitem{Degollado:2018ypf}
Juan~Carlos Degollado, Carlos A.~R. Herdeiro, and Eugen Radu.
\newblock {Effective stability against superradiance of Kerr black holes with
  synchronised hair}.
\newblock {\em Phys. Lett. B}, 781:651--655, 2018.

\bibitem{Nakamura:1987zz}
T.~Nakamura, K.~Oohara, and Y.~Kojima.
\newblock {General Relativistic Collapse to Black Holes and Gravitational Waves
  from Black Holes}.
\newblock {\em Prog. Theor. Phys. Suppl.}, 90:1--218, 1987.

\bibitem{Whiting:1988vc}
Bernard~F. Whiting.
\newblock {Mode Stability of the Kerr Black Hole}.
\newblock {\em J. Math. Phys.}, 30:1301, 1989.

\bibitem{Penrose:1964wq}
Roger Penrose.
\newblock {Gravitational collapse and space-time singularities}.
\newblock {\em Phys. Rev. Lett.}, 14:57--59, 1965.

\bibitem{LIGOScientific:2020iuh}
R.~Abbott et~al.
\newblock {GW190521: A Binary Black Hole Merger with a Total Mass of $150
  M_{\odot}$}.
\newblock {\em Phys. Rev. Lett.}, 125(10):101102, 2020.

\bibitem{LIGOScientific:2020ufj}
R.~Abbott et~al.
\newblock {Properties and Astrophysical Implications of the 150 M$_\odot$
  Binary Black Hole Merger GW190521}.
\newblock {\em Astrophys. J. Lett.}, 900(1):L13, 2020.

\bibitem{Falcke:1999pj}
Heino Falcke, Fulvio Melia, and Eric Agol.
\newblock {Viewing the shadow of the black hole at the galactic center}.
\newblock {\em Astrophys. J. Lett.}, 528:L13, 2000.

\bibitem{Cunha:2018acu}
Pedro V.~P. Cunha and Carlos A.~R. Herdeiro.
\newblock {Shadows and strong gravitational lensing: a brief review}.
\newblock {\em Gen. Rel. Grav.}, 50(4):42, 2018.

\bibitem{Carter:1971zc}
B.~Carter.
\newblock {Axisymmetric Black Hole Has Only Two Degrees of Freedom}.
\newblock {\em Phys. Rev. Lett.}, 26:331--333, 1971.

\bibitem{Robinson:1975bv}
D.~C. Robinson.
\newblock {Uniqueness of the Kerr black hole}.
\newblock {\em Phys. Rev. Lett.}, 34:905--906, 1975.

\bibitem{Chrusciel:2012jk}
Piotr~T. Chrusciel, Joao Lopes~Costa, and Markus Heusler.
\newblock {Stationary Black Holes: Uniqueness and Beyond}.
\newblock {\em Living Rev. Rel.}, 15:7, 2012.

\bibitem{Ruffini:1971bza}
Remo Ruffini and John~A. Wheeler.
\newblock {Introducing the black hole}.
\newblock {\em Phys. Today}, 24(1):30, 1971.

\bibitem{Herdeiro:2015waa}
Carlos A.~R. Herdeiro and Eugen Radu.
\newblock {Asymptotically flat black holes with scalar hair: a review}.
\newblock {\em Int. J. Mod. Phys. D}, 24(09):1542014, 2015.

\bibitem{Volkov:2016ehx}
Mikhail~S. Volkov.
\newblock {Hairy black holes in the XX-th and XXI-st centuries}.
\newblock In {\em {14th Marcel Grossmann Meeting on Recent Developments in
  Theoretical and Experimental General Relativity, Astrophysics, and
  Relativistic Field Theories}}, volume~2, pages 1779--1798, 2017.

\bibitem{Bekenstein:1972ny}
J.~D. Bekenstein.
\newblock {Transcendence of the law of baryon-number conservation in black hole
  physics}.
\newblock {\em Phys. Rev. Lett.}, 28:452--455, 1972.

\bibitem{Hui:2016ltb}
Lam Hui, Jeremiah~P. Ostriker, Scott Tremaine, and Edward Witten.
\newblock {Ultralight scalars as cosmological dark matter}.
\newblock {\em Phys. Rev. D}, 95(4):043541, 2017.

\bibitem{Suarez:2013iw}
Abril Su\'arez, Victor~H. Robles, and Tonatiuh Matos.
\newblock {A Review on the Scalar Field/Bose-Einstein Condensate Dark Matter
  Model}.
\newblock {\em Astrophys. Space Sci. Proc.}, 38:107--142, 2014.

\bibitem{Freitas:2021cfi}
Felipe~F. Freitas, Carlos A.~R. Herdeiro, Ant\'onio~P. Morais, Ant\'onio
  Onofre, Roman Pasechnik, Eugen Radu, Nicolas Sanchis-Gual, and Rui Santos.
\newblock {Ultralight bosons for strong gravity applications from simple
  Standard Model extensions}.
\newblock {\em JCAP}, 12(12):047, 2021.

\bibitem{Brito:2015oca}
Richard Brito, Vitor Cardoso, and Paolo Pani.
\newblock {Superradiance}: {New Frontiers in Black Hole Physics}.
\newblock {\em Lect. Notes Phys.}, 906:pp.1--237, 2015.

\bibitem{East:2017ovw}
William~E. East and Frans Pretorius.
\newblock {Superradiant Instability and Backreaction of Massive Vector Fields
  around Kerr Black Holes}.
\newblock {\em Phys. Rev. Lett.}, 119(4):041101, 2017.

\bibitem{Herdeiro:2017phl}
Carlos A.~R. Herdeiro and Eugen Radu.
\newblock {Dynamical Formation of Kerr Black Holes with Synchronized Hair: An
  Analytic Model}.
\newblock {\em Phys. Rev. Lett.}, 119(26):261101, 2017.

\bibitem{Herdeiro:2014goa}
Carlos A.~R. Herdeiro and Eugen Radu.
\newblock {Kerr black holes with scalar hair}.
\newblock {\em Phys. Rev. Lett.}, 112:221101, 2014.

\bibitem{Herdeiro:2016tmi}
Carlos Herdeiro, Eugen Radu, and Helgi R\'unarsson.
\newblock {Kerr black holes with Proca hair}.
\newblock {\em Class. Quant. Grav.}, 33(15):154001, 2016.

\bibitem{Doneva:2017bvd}
Daniela~D. Doneva and Stoytcho~S. Yazadjiev.
\newblock {New Gauss-Bonnet Black Holes with Curvature-Induced Scalarization in
  Extended Scalar-Tensor Theories}.
\newblock {\em Phys. Rev. Lett.}, 120(13):131103, 2018.

\bibitem{Silva:2017uqg}
Hector~O. Silva, Jeremy Sakstein, Leonardo Gualtieri, Thomas~P. Sotiriou, and
  Emanuele Berti.
\newblock {Spontaneous scalarization of black holes and compact stars from a
  Gauss-Bonnet coupling}.
\newblock {\em Phys. Rev. Lett.}, 120(13):131104, 2018.

\bibitem{Damour:1993hw}
Thibault Damour and Gilles Esposito-Farese.
\newblock {Nonperturbative strong field effects in tensor - scalar theories of
  gravitation}.
\newblock {\em Phys. Rev. Lett.}, 70:2220--2223, 1993.

\bibitem{Kaup:1968zz}
David~J. Kaup.
\newblock {Klein-Gordon Geon}.
\newblock {\em Phys. Rev.}, 172:1331--1342, 1968.

\bibitem{Ruffini:1969qy}
Remo Ruffini and Silvano Bonazzola.
\newblock {Systems of selfgravitating particles in general relativity and the
  concept of an equation of state}.
\newblock {\em Phys. Rev.}, 187:1767--1783, 1969.

\bibitem{Schunck:2003kk}
Franz~E. Schunck and Eckehard~W. Mielke.
\newblock {General relativistic boson stars}.
\newblock {\em Class. Quant. Grav.}, 20:R301--R356, 2003.

\bibitem{Brito:2015pxa}
Richard Brito, Vitor Cardoso, Carlos A.~R. Herdeiro, and Eugen Radu.
\newblock {Proca stars: Gravitating Bose\textendash{}Einstein condensates of
  massive spin 1 particles}.
\newblock {\em Phys. Lett. B}, 752:291--295, 2016.

\bibitem{Herdeiro:2017fhv}
Carlos A.~R. Herdeiro, Alexandre~M. Pombo, and Eugen Radu.
\newblock {Asymptotically flat scalar, Dirac and Proca stars: discrete vs.
  continuous families of solutions}.
\newblock {\em Phys. Lett. B}, 773:654--662, 2017.

\bibitem{Herdeiro:2019mbz}
C.~Herdeiro, I.~Perapechka, E.~Radu, and Ya. Shnir.
\newblock {Asymptotically flat spinning scalar, Dirac and Proca stars}.
\newblock {\em Phys. Lett. B}, 797:134845, 2019.

\bibitem{Liebling:2012fv}
Steven~L. Liebling and Carlos Palenzuela.
\newblock {Dynamical Boson Stars}.
\newblock {\em Living Rev. Rel.}, 15:6, 2012.

\bibitem{Sanchis-Gual:2019ljs}
N.~Sanchis-Gual, F.~Di~Giovanni, M.~Zilh\~ao, C.~Herdeiro, P.~Cerd\'a-Dur\'an,
  J.~A. Font, and E.~Radu.
\newblock {Nonlinear Dynamics of Spinning Bosonic Stars: Formation and
  Stability}.
\newblock {\em Phys. Rev. Lett.}, 123(22):221101, 2019.

\bibitem{Seidel:1993zk}
Edward Seidel and Wai-Mo Suen.
\newblock {Formation of solitonic stars through gravitational cooling}.
\newblock {\em Phys. Rev. Lett.}, 72:2516--2519, 1994.

\bibitem{Guzman:2006yc}
F.~Siddhartha Guzman and L.~Arturo Urena-Lopez.
\newblock {Gravitational cooling of self-gravitating Bose-Condensates}.
\newblock {\em Astrophys. J.}, 645:814--819, 2006.

\bibitem{DiGiovanni:2018bvo}
Fabrizio Di~Giovanni, Nicolas Sanchis-Gual, Carlos A.~R. Herdeiro, and
  Jos\'e~A. Font.
\newblock {Dynamical formation of Proca stars and quasistationary solitonic
  objects}.
\newblock {\em Phys. Rev. D}, 98(6):064044, 2018.

\bibitem{strogaatz}
S.~Strogatz.
\newblock {\em Sync: How Order Emerges from Chaos in the Universe, Nature, and
  Daily Life}.
\newblock Hachette Books, New York, 2004.

\bibitem{Hut}
P.~{Hut}.
\newblock {Tidal evolution in close binary systems.}
\newblock {\em Astron. \& Astrophys.}, 99:126--140, June 1981.

\bibitem{Hod:2012px}
Shahar Hod.
\newblock {Stationary Scalar Clouds Around Rotating Black Holes}.
\newblock {\em Phys. Rev. D}, 86:104026, 2012.
\newblock [Erratum: Phys.Rev.D 86, 129902 (2012)].

\bibitem{Dias:2011at}
Oscar J.~C. Dias, Gary~T. Horowitz, and Jorge~E. Santos.
\newblock {Black holes with only one Killing field}.
\newblock {\em JHEP}, 07:115, 2011.

\bibitem{Herdeiro:2021znw}
Carlos A.~R. Herdeiro, Eugen Radu, and Nuno~M. Santos.
\newblock {A bound on energy extraction (and hairiness) from superradiance}.
\newblock {\em Phys. Lett. B}, 824:136835, 2022.

\bibitem{Dolan:2012yt}
Sam~R. Dolan.
\newblock {Superradiant instabilities of rotating black holes in the time
  domain}.
\newblock {\em Phys. Rev. D}, 87(12):124026, 2013.

\bibitem{Sanchis-Gual:2020mzb}
Nicolas Sanchis-Gual, Miguel Zilh\~ao, Carlos Herdeiro, Fabrizio Di~Giovanni,
  Jos\'e~A. Font, and Eugen Radu.
\newblock {Synchronized gravitational atoms from mergers of bosonic stars}.
\newblock {\em Phys. Rev. D}, 102(10):101504, 2020.

\bibitem{Collodel:2021jwi}
Lucas~G. Collodel, Daniela~D. Doneva, and Stoytcho~S. Yazadjiev.
\newblock {Equatorial EMRIs in KBHsSH Spacetimes}.
\newblock 8 2021.

\bibitem{Cunha:2015yba}
Pedro V.~P. Cunha, Carlos A.~R. Herdeiro, Eugen Radu, and Helgi~F. Runarsson.
\newblock {Shadows of Kerr black holes with scalar hair}.
\newblock {\em Phys. Rev. Lett.}, 115(21):211102, 2015.

\bibitem{Cunha:2019ikd}
Pedro V.~P. Cunha, Carlos A.~R. Herdeiro, and Eugen Radu.
\newblock {EHT constraint on the ultralight scalar hair of the M87 supermassive
  black hole}.
\newblock {\em Universe}, 5(12):220, 2019.

\bibitem{Cunha:2016bpi}
Pedro V.~P. Cunha, Carlos A.~R. Herdeiro, Eugen Radu, and Helgi~F. Runarsson.
\newblock {Shadows of Kerr black holes with and without scalar hair}.
\newblock {\em Int. J. Mod. Phys. D}, 25(09):1641021, 2016.

\bibitem{Kanti:1995vq}
P.~Kanti, N.~E. Mavromatos, J.~Rizos, K.~Tamvakis, and E.~Winstanley.
\newblock {Dilatonic black holes in higher curvature string gravity}.
\newblock {\em Phys. Rev. D}, 54:5049--5058, 1996.

\bibitem{Kleihaus:2015aje}
Burkhard Kleihaus, Jutta Kunz, Sindy Mojica, and Eugen Radu.
\newblock {Spinning black holes in
  Einstein\textendash{}Gauss-Bonnet\textendash{}dilaton theory: Nonperturbative
  solutions}.
\newblock {\em Phys. Rev. D}, 93(4):044047, 2016.

\bibitem{Sotiriou:2014pfa}
Thomas~P. Sotiriou and Shuang-Yong Zhou.
\newblock {Black hole hair in generalized scalar-tensor gravity: An explicit
  example}.
\newblock {\em Phys. Rev. D}, 90:124063, 2014.

\bibitem{Delgado:2020rev}
Jorge F.~M. Delgado, Carlos A.~R. Herdeiro, and Eugen Radu.
\newblock {Spinning black holes in shift-symmetric Horndeski theory}.
\newblock {\em JHEP}, 04:180, 2020.

\bibitem{Benkel:2016kcq}
Robert Benkel, Thomas~P. Sotiriou, and Helvi Witek.
\newblock {Dynamical scalar hair formation around a Schwarzschild black hole}.
\newblock {\em Phys. Rev. D}, 94(12):121503, 2016.

\bibitem{Antoniou:2017acq}
G.~Antoniou, A.~Bakopoulos, and P.~Kanti.
\newblock {Evasion of No-Hair Theorems and Novel Black-Hole Solutions in
  Gauss-Bonnet Theories}.
\newblock {\em Phys. Rev. Lett.}, 120(13):131102, 2018.

\bibitem{Cunha:2019dwb}
Pedro V.~P. Cunha, Carlos A.~R. Herdeiro, and Eugen Radu.
\newblock {Spontaneously Scalarized Kerr Black Holes in Extended
  Scalar-Tensor\textendash{}Gauss-Bonnet Gravity}.
\newblock {\em Phys. Rev. Lett.}, 123(1):011101, 2019.

\bibitem{Doneva:2021dqn}
Daniela~D. Doneva and Stoytcho~S. Yazadjiev.
\newblock {Dynamics of the nonrotating and rotating black hole scalarization}.
\newblock {\em Phys. Rev. D}, 103(6):064024, 2021.

\bibitem{Herdeiro:2018wub}
Carlos A.~R. Herdeiro, Eugen Radu, Nicolas Sanchis-Gual, and Jos\'e~A. Font.
\newblock {Spontaneous Scalarization of Charged Black Holes}.
\newblock {\em Phys. Rev. Lett.}, 121(10):101102, 2018.

\bibitem{Collodel:2019kkx}
Lucas~G. Collodel, Burkhard Kleihaus, Jutta Kunz, and Emanuele Berti.
\newblock {Spinning and excited black holes in
  Einstein-scalar-Gauss\textendash{}Bonnet theory}.
\newblock {\em Class. Quant. Grav.}, 37(7):075018, 2020.

\bibitem{Blazquez-Salcedo:2018jnn}
Jose~Luis Bl\'azquez-Salcedo, Daniela~D. Doneva, Jutta Kunz, and Stoytcho~S.
  Yazadjiev.
\newblock {Radial perturbations of the scalarized Einstein-Gauss-Bonnet black
  holes}.
\newblock {\em Phys. Rev. D}, 98(8):084011, 2018.

\bibitem{Dima:2020yac}
Alexandru Dima, Enrico Barausse, Nicola Franchini, and Thomas~P. Sotiriou.
\newblock {Spin-induced black hole spontaneous scalarization}.
\newblock {\em Phys. Rev. Lett.}, 125(23):231101, 2020.

\bibitem{Herdeiro:2020wei}
Carlos A.~R. Herdeiro, Eugen Radu, Hector~O. Silva, Thomas~P. Sotiriou, and
  Nicol\'as Yunes.
\newblock {Spin-induced scalarized black holes}.
\newblock {\em Phys. Rev. Lett.}, 126(1):011103, 2021.

\bibitem{Berti:2020kgk}
Emanuele Berti, Lucas~G. Collodel, Burkhard Kleihaus, and Jutta Kunz.
\newblock {Spin-induced black-hole scalarization in
  Einstein-scalar-Gauss-Bonnet theory}.
\newblock {\em Phys. Rev. Lett.}, 126(1):011104, 2021.

\bibitem{Cardoso:2016rao}
Vitor Cardoso, Edgardo Franzin, and Paolo Pani.
\newblock {Is the gravitational-wave ringdown a probe of the event horizon?}
\newblock {\em Phys. Rev. Lett.}, 116(17):171101, 2016.
\newblock [Erratum: Phys.Rev.Lett. 117, 089902 (2016)].

\bibitem{Cunha:2017eoe}
Pedro V.~P. Cunha, Carlos A.~R. Herdeiro, and Eugen Radu.
\newblock {Fundamental photon orbits: black hole shadows and spacetime
  instabilities}.
\newblock {\em Phys. Rev. D}, 96(2):024039, 2017.

\bibitem{Cunha:2017qtt}
Pedro V.~P. Cunha, Emanuele Berti, and Carlos A.~R. Herdeiro.
\newblock {Light-Ring Stability for Ultracompact Objects}.
\newblock {\em Phys. Rev. Lett.}, 119(25):251102, 2017.

\bibitem{Keir:2014oka}
Joe Keir.
\newblock {Slowly decaying waves on spherically symmetric spacetimes and
  ultracompact neutron stars}.
\newblock {\em Class. Quant. Grav.}, 33(13):135009, 2016.

\bibitem{Bustillo:2020syj}
Juan~Calder\'on Bustillo, Nicolas Sanchis-Gual, Alejandro Torres-Forn\'e,
  Jos\'e~A. Font, Avi Vajpeyi, Rory Smith, Carlos Herdeiro, Eugen Radu, and
  Samson H.~W. Leong.
\newblock {GW190521 as a Merger of Proca Stars: A Potential New Vector Boson of
  $8.7\times 10^{-13}$ eV}.
\newblock {\em Phys. Rev. Lett.}, 126(8):081101, 2021.

\bibitem{Olivares:2018abq}
Hector Olivares, Ziri Younsi, Christian~M. Fromm, Mariafelicia De~Laurentis,
  Oliver Porth, Yosuke Mizuno, Heino Falcke, Michael Kramer, and Luciano
  Rezzolla.
\newblock {How to tell an accreting boson star from a black hole}.
\newblock {\em Mon. Not. Roy. Astron. Soc.}, 497(1):521--535, 2020.

\bibitem{Herdeiro:2021lwl}
Carlos A.~R. Herdeiro, Alexandre~M. Pombo, Eugen Radu, Pedro V.~P. Cunha, and
  Nicolas Sanchis-Gual.
\newblock {The imitation game: Proca stars that can mimic the Schwarzschild
  shadow}.
\newblock {\em JCAP}, 04:051, 2021.

\end{thebibliography}
\end{document}